\documentstyle[twoside, epsf]{article}

\input ibvs2.sty

\begin{document}
\alph{mpfootnote}
\IBVShead{5xxx}{00 Month 200x}

\IBVStitle{Rapid Variations in V2275 Cyg (Nova Cyg 2001\#2)}

\IBVSauth{Garnavich, P.M.$^{1,2}$, Macdonald,  A.J.$^1$, Wu, B.$^{3}$, Easterday, S.M.$^1$,
Libal, A.J.$^1$, Palumbo, A.$^1$, Quinn, M.A.$^1$}

\IBVSinst{Department of Physics, University of Notre Dame, Notre Dame, IN 46566}
\IBVSinst{email: pgarnavi@nd.edu}
\IBVSinst{Department of Electrical Engineering, University of Notre Dame, Notre Dame, IN 46566 }

\SIMBADobjAlias{V2275~Cyg}{Nova Cygni 2001\#2}
\IBVStyp{CWA}
\IBVSkey{photometry}
\IBVSabs{We present time-resolved CCD photometry of V2275~Cyg 790 days after the nova outburst.}
\IBVSabs{The data show a 20\% peak-to-peak oscillation with a period of 24 minutes.}
\IBVSabs{The period appears stable over the two nights of observation, but QPO can not be ruled out.}

\begintext

Nova Cygni 2001\#2 was discovered by A. Tago and K. Matayama on Aug. 18 (Nakamura 2001)
at a magnitude of 6.6 .  The brightness decay from maximum was one of the fastest
ever recorded and it shows characteristics of recurrent novae (Kiss et al. 2002). Time
resolved photometry in 2002 Oct. by Balman et al. (2003) revealed large amplitude variations
with a period of 8 or 11 hours which might be associated with its orbital period.

\IBVSfig{10cm}{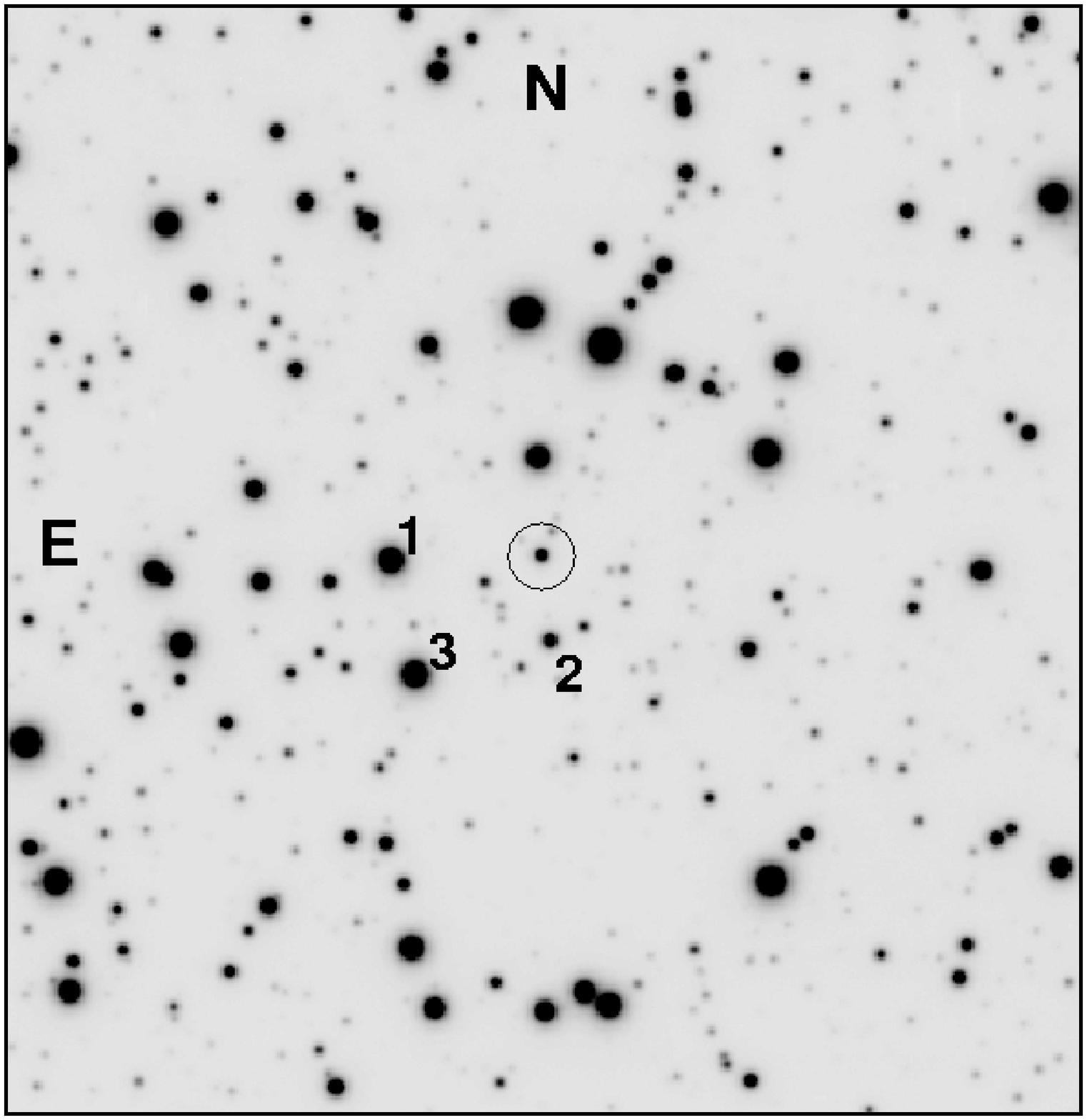}{
A finder chart for V2275 Cygni (circled) and comparison stars (listed in Table~1). The image
is an average of 162 $V$-band exposures, each 120 seconds in length. The field of view is $3.5'$
on a side. }

We observed V2275~Cyg beginning on HJD2452930.61 (2003 Oct 18) with the 1.8-m Vatican Advanced
Technology Telescope (VATT). This was 790 days after maximum light. Observations continued
the next night starting at HJD2452931.59.
The CCD was binned 2$\times$2 providing a scale
of 0.4~arcsec/pixel. $V$-band exposures were 120~seconds with 30 second readout time covering
about 10 hours over the two nights. The resulting images were bias subtracted and flat-fielded
and instrumental magnitudes were measured using aperture photometry. Three comparison
stars near the nova (Figure~1) were also measured and their positions listed in Table~1.
Approximate
$V$-band magnitudes for the stars were estimated from zero-point and airmass coefficients
measured at the VATT on earlier runs with  standard magnitude errors estimated to be
$\pm 0.05$~mag. 

\IBVSfig{10cm}{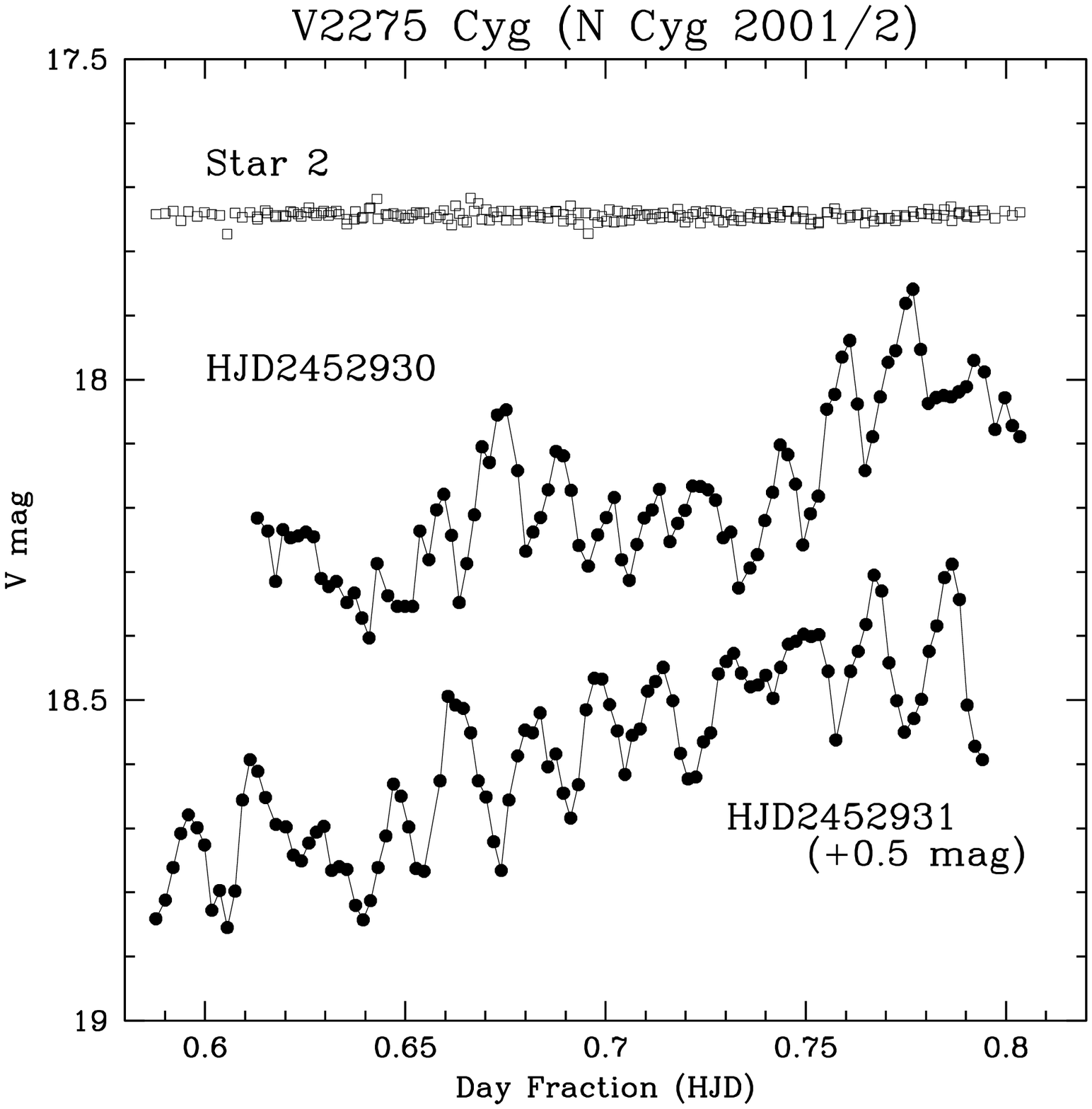}{
The $V$-band light curves of V2275~Cyg obtained with the VATT. The Oct. 19 data have
been displaced by $+0.5$~mag.  } 

The VATT light curves are shown in Figure~2 and were obtained by subtracting the
instrumental magnitude of Star~1 from the nova instrumental magnitude.  The light curve for Star~2
relative to Star~1 is also shown and demonstrates both stars were constant over the observing
run. Variations in individual measurements show an RMS scatter of 0.01~mag per
exposure for $V\sim 18$~mag stars.

V2275~Cyg clearly shows light variations with a full amplitude of 0.2~mag and a period of
about 20 minutes as well as a slower brightening trend on both nights. At the time of
the observations V2275~Cyg varied between $18.0<V<18.5$ mag. The long-period
variation suggests a period $>7$ hours and may be the same phenomenon seen by Balman et al. (2003).
The short period variation has not been previously observed and may a quasi-periodic oscillation (QPO)
or a stable periodicity such as a spinning white dwarf. Power-spectrum analysis gives a period
of 0.410$\pm 0.005$ hours for the variation. A plot of this period against the normalized light
curve is shown in Figure~3 and indicates that this period was stable over the two nights
of data. This supports the possibility that the short-term light variations come from
reprocessing of light from a asynchronous spinning white dwarf, but more data is
needed to confirm the stability of the period.

We acknowledge travel assistance from the University of Notre Dame Department of Physics.
This work based on
observations with the VATT: the Alice P. Lennon Telescope and the Thomas J. Bannan
Astrophysics Facility.

\IBVSfig{10cm}{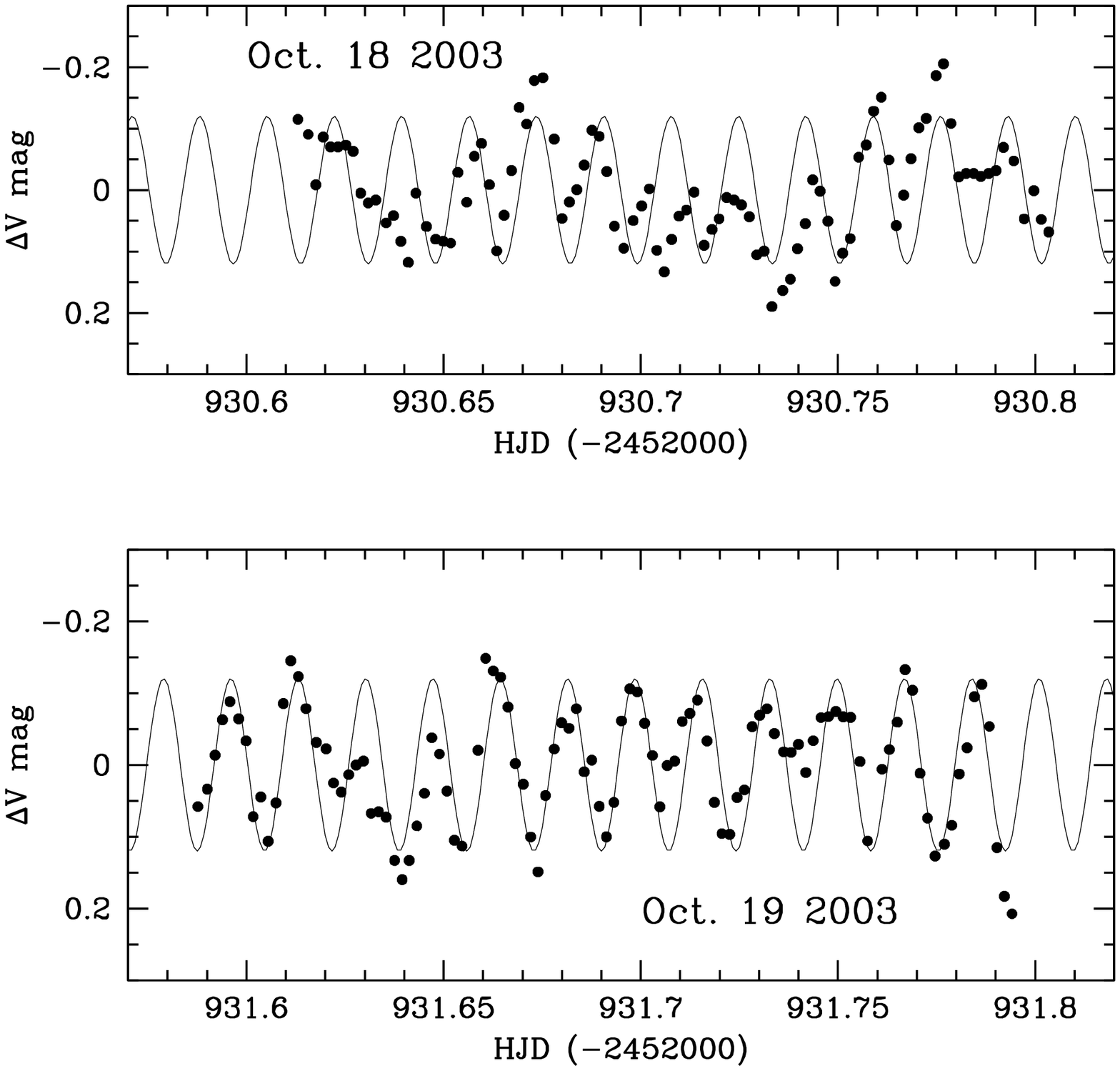}{The normalized light curve compared to a sinusoidal oscillation
with a period of 0.41 hours.  }

\begin{center}
{Table 1. Comparison Stars \\}
\vskip 2mm
\begin{tabular}{lccc}
\hline
Star   &  RA (2000)  &  Dec (2000) & V mag   \\
\hline
1 & 21:03:04.80 & +48:45:51  & 15.50  \\
2 & 21:03:01.96 & +48:45:37  & 17.75  \\
3 & 21:03:04.38 & +48:45:31  & 15.33  \\
\hline
\end{tabular}
\end{center}

\references

Balman, S., et al.  2003, {\it IAU Circ}, {8074}

Kiss, L.L., et al. 2002, {\it A\&A}, {\bf 384}, 982

Nakamura, A. 2001, {\it IAU Circ}, {7686}

\endreferences

\IBVSedata{5xxx-t2.txt}
\IBVSedata{5xxx-t3.txt}

\IBVSefigure{5xxx-f2.ps}
\IBVSefigure{5xxx-f3.ps}
\IBVSefigure{5xxx-f4.ps}
\IBVSefigure{5xxx-f5.ps}

\end{document}